\documentclass[runningheads]{llncs}
\usepackage[T1]{fontenc}
\usepackage{graphicx,amsmath,amssymb,mathrsfs,framed,esint,slashed,upgreek}
\usepackage[colorlinks]{hyperref}
\usepackage{color}

\newcommand{\rem}[1]{}

\newcommand{\de}{{\rm d}}

\newcommand{\bX}{{\mathbf{X}}}

\newcommand{\bzeta}{{\boldsymbol{\zeta}}}

\newcommand{\bz}{{\mathbf{z}}}

\newcommand{\bA}{{\boldsymbol{\cal A}}}

\newcommand{\beq}{\begin{equation}}
\newcommand{\eeq}{\end{equation}}
\newcommand{\ben}{\begin{eqnarray}}
\newcommand{\een}{\end{eqnarray}}

\begin{document}
\title{Lagrangian trajectories and closure models\\in mixed quantum-classical dynamics\thanks{This work was made possible through the support of Grant 62210 from the John Templeton Foundation. The opinions expressed in this publication are those of the authors and do not necessarily reflect the views of the John Templeton Foundation.}}
\titlerunning{Mixed quantum-classical dynamics}
%
\author{
\makebox{Cesare Tronci\inst{1,2}
\orcidID{0000-0002-8868-8027} 
\and Fran\c{c}ois Gay-Balmaz\inst{3}
\orcidID{0000-0001-9411-1382}
}}
\authorrunning{F. Gay-Balmaz and C. Tronci}
%
\institute{$^1$Department of Mathematics, University of Surrey, Guildford, UK\\
\makebox{$^2$Department of Physics and Engineering Physics, Tulane University, New Orleans, USA}\\
 \email{c.tronci@surrey.ac.uk}\\
 \medskip
\makebox{\hspace{-.35cm}$^3$CNRS \& Laboratoire de M\'et\'eorologie Dynamique, \'Ecole Normale Sup\'erieure, Paris, France}\\
\email{francois.gay-balmaz@lmd.ens.fr}}

\maketitle              
\begin{abstract}
Mixed quantum-classical models have been proposed in  several contexts to overcome the computational challenges of fully quantum approaches. However, current models   typically suffer from long-standing consistency issues, and, in some cases, invalidate Heisenberg's uncertainty principle. Here, we present a fully Hamiltonian theory of quantum-classical dynamics that appears to be the first to ensure a series of consistency properties, beyond positivity of quantum and classical densities. Based on Lagrangian phase-space paths, the model  possesses a quantum-classical Poincar\'e integral invariant as well as infinite classes of Casimir functionals. We also exploit Lagrangian trajectories to formulate a finite-dimensional closure scheme for numerical implementations.

\end{abstract}
\section{Introduction}

The search for a mixed quantum-classical description of many-body quantum systems is motivated by the formidable challenges posed by the curse of dimensionality appearing in fully quantum approaches. For example, it is common practice in molecular dynamics  to approximate nuclei as classical particles while retaining a fully quantum electronic description. Similar mixed quantum-classical approximations have also been proposed in quantum plasmas and, more recently, in magnon spintronics.

\paragraph{Hybrid quantum-classical models.} Despite the computational appeal, the interaction dynamics of quantum and classical degrees of freedom continues to represent a challenging question since the currently  available models suffer from several consistency issues. In some cases, the Heisenberg principle is lost due to the fact that the quantum density matrix is allowed to change its sign. In some other cases, the model does not reduce to uncoupled quantum and classical dynamics in the absence of a quantum-classical interaction potential. 
At a computational level, the most popular approach is probably  the Ehrenfest model, which reads
\beq\label{Ehrenfest}
\partial_t D+\operatorname{div}\!\big(D\langle\bX_{\widehat{H}}\rangle\big)=0
,\qquad\qquad 
i\hbar\big(\partial_t\psi+\langle\bX_{\widehat{H}}\rangle\cdot\nabla\psi\big)=\widehat{H}\psi,
\eeq
where $\bX_{\widehat{H}}=\big(\partial_p\widehat{H},-\partial_q\widehat{H}\big)$.
Here, $D(q,p)$ is the classical density, while $\psi(x;q,p)$ is a wavefunction depending on the quantum $x-$coordinate and parameterized by the classical coordinates $(q,p)$. Also,  $\widehat{H}(q,p)$ is a quantum Hamiltonian operator depending on $(q,p)$ and we have resorted to the usual notation $\langle \widehat{A}\rangle=\langle\psi|\widehat{A}(q,p)\psi\rangle$, where $\langle\psi_1|\psi_2\rangle=\int\psi^*_1(x)\psi_2(x)\,\de x$. In this setting, the matrix elements of the quantum density operator are given  as 
\[
\hat\rho(x,x')=\int D(q,p)\psi(x;q,p)\psi^*(x';q,p)\,\de q\de p.
\]
 Despite its wide popularity, the Ehrenfest model \eqref{Ehrenfest} fails to reproduce realistic levels of decoherence, which is usually expressed in terms of the norm squared $\|\hat\rho\|^2$ of the density operator, a quantity also known as \emph{quantum purity}. 

Any quantum-classical description beyond the Ehrenfest model must still ensure its five  consistency properties: 1) the classical system is identified by a phase-space probability density at all times; 2) the quantum system is identified by a positive-semidefinite density operator $\hat\rho$ at all times; 3) the model is covariant under both quantum unitary transformations and classical canonical transformations; 4) in the absence of an interaction potential, the model reduces to uncoupled quantum and classical dynamics; 5) in the presence of an interaction potential, the {\it quantum purity} $\|\hat\rho\|^2$ is not a constant of motion (decoherence property). A model satisfying properties 1)-4), but not 5) is the \emph{mean-field model}
\beq\label{MFmod}
\frac{\partial D}{\partial t}+\big\{D,{\langle\widehat{H}\rangle}\big\}=0
,\qquad\qquad 
i\hbar\frac{\de\hat\rho}{\de t}=\bigg[\int \!D\widehat{H}\,\de q\de p,\hat\rho\bigg]
\,,
\eeq
where $\{\cdot,\cdot\}$ denotes the canonical Poisson bracket. Here, we notice that  the quantum density matrix $\hat\rho$ does not carry any dependence on the phase-space coordinates. Most recent efforts in quantum-classical methods are addressed to the design of new models beyond the Ehrenfest system that can better capture decoherence effects and still retain all the  consistency properties above.

\paragraph{Beyond the Ehrenfest model.}  
Blending Koopman wavefunctions in classical mechanics  with the geometry of prequantum theory, we recently formulated a quantum-classical model \cite{GBTr22,GBTr21} which was developed in two stages. First, we provided an early quantum-classical model \cite{BoGBTr19} that succeeded in satisfying only the properties 2)-5). 
Then, more recently, we upgraded this model in such a way that property 1) is also secured \cite{GBTr22,GBTr21}. This  upgrade was achieved by combining Lagrangian trajectories on the classical phase-space with  a gauge principle which ensures that classical phases are \emph{unobservable}, that is they do not contribute to measurable expectation values. 
Inspired by Sudarshan's work \cite{Sudarshan}, this combination  leads naturally to crucial properties such as the characterization of entropy functionals and the Poincar\'e integral invariant in the context of quantum-classical dynamics. Nevertheless, the model is nonlinear and its explicit form is rather intricate due to the appearance of the non-Abelian gauge connection $i[P,\nabla P]$, where $P(x,x';q,p)=\psi(x;q,p)\psi^*(x';q,p)$. In particular, this gauge connection  emerges through the (Hermitian) operator-valued vector field  $\boldsymbol{\widehat{\Gamma}}={i}[P,\bX_ P]$ in such a way that  the model proposed in \cite{GBTr22,GBTr21} reads 
\beq
\partial_t D+\operatorname{div}(D\boldsymbol{\cal X})=0
,\qquad\qquad
i\hbar(\partial_t\psi+\boldsymbol{\cal X}\cdot\nabla\psi)=\widehat{\cal H}\psi,
\label{HybEq1}
\eeq
with
\beq
\boldsymbol{\cal X}=
\langle\bX_{\widehat{H}}\rangle+\frac\hbar{2D}\operatorname{Tr}\!\big( \bX_{{\widehat{H}}}\cdot\nabla (D\boldsymbol{\widehat{\Gamma}})-(D\boldsymbol{\widehat{\Gamma}})\cdot\nabla\bX_{{\widehat{H}}}
\big),
\label{HybEq2}
\eeq
and
\beq
\widehat{\mathcal{H}}= \widehat{H}+i\hbar\Big(\{P,{\widehat{H}}\}+\{\widehat{H},P\}-\frac{1}{2D}[\{ D,\widehat{H}\},P] \Big).
\label{HybEq3}
\eeq  
Thus, we conclude that the vector field $\boldsymbol{\cal X}$ and the Hermitian generator $\widehat{\mathcal{H}}$ can be regarded as $\hbar-$modifications of the original Ehrenfest quantities $\langle\bX_{\widehat{H}}\rangle$ and $\widehat{H}$, respectively.  While equations \eqref{HybEq1}-\eqref{HybEq3} appear hardly tractable at first sight, a direct calculation of $\operatorname{div}\!\boldsymbol{\cal X}$ reveals that no gradients of order higher than two appear in the equations \eqref{HybEq1}. In addition,  the Hamiltonian/variational structure of this system unfolds much of the features occurring in quantum-classical coupling. Thus, we consider the equations above as a platform for the formulation of simplified closure models that can be used in physically relevant cases.

\paragraph{Trajectory-based numerical algorithms.} The presence of transport terms in equations \eqref{HybEq1} results from the predominant role played by Lagrangian trajectories on the classical phase-space. These terms hint to the possibility of using characteristic curves to design trajectory-based schemes for mixed quantum-classical simulation codes in molecular dynamics \cite{HoRaTr21}. However, the presence of several gradients in the expression of the transport vector field $\boldsymbol{\cal X}$ prevents the direct application of trajectory-based methods, which instead can be readily used for the Ehrenfest equations \eqref{Ehrenfest}. In the latter case, if we denote ${\bz=(q,p)}$, we observe that the first equation is solved by $D(\bz,t)=\sum_{a=1}^Nw_a\delta(\bz-\bzeta_a(t))$ with $\dot{\bzeta}_a=\langle\bX_{\widehat{H}}\rangle|_{\bz=\bzeta_a}$. Here, the quantity $\langle\bX_{\widehat{H}}\rangle|_{\bz=\bzeta_a}$ requires evaluating ${\psi_a(t):=\psi(\bzeta_a(t),t)}$ at all times and this can indeed be done by multiplying the second in \eqref{Ehrenfest} by $D$ and then integrating, so that $i\hbar\dot{\psi}_a(t)=\widehat{H}(\bzeta_a(t))\psi_a(t)$. Eventually, direct application of the trajectory method to the Ehrenfest model leads to the  equations
\beq
\dot{q}_a=\partial_{p_a\!}\langle\psi_a|\widehat{H}_a\psi_a\rangle
,\qquad\ 
\dot{p}_a=-\partial_{q_a\!}\langle\psi_a|\widehat{H}_a\psi_a\rangle
,\qquad\ 
i\hbar\dot{\psi}_a=\widehat{H}_a\psi_a
,
\label{MFeqs}
\eeq
where ${\widehat{H}_a:=\widehat{H}(q_a,p_a)}$ and $\bzeta_a=(q_a,p_a)$. Then, for a finite-dimensional quantum Hilbert space, $\psi(\bz,t)\in\Bbb{C}^n$ and  the quantum density matrix is  $\hat\rho=\int D\psi\psi^\dagger\de^2z=\sum_a w_a\psi_a\psi_a^\dagger$. 

The same approach may be applied to the mean-field model by writing  $\hat\rho=\sum_{i=1}^Nw_a\psi_a\psi_a^\dagger$ and $D=\sum_{i=1}^Nw_a\delta(\bz-\bzeta_a(t))$ in \eqref{MFmod}. Then,
in the case $N=1$ (only one trajectory) the closure of the Ehrenfest equations coincides with the closure of the mean-field model for the interaction of a classical particle with a pure quantum state. Notice, however, that the equations \eqref{MFeqs}  generally account for decoherence effects when  ${N>1}$, while the same is not true for the mean-field model. Thus, while the Ehrenfest  and the mean-field models are regarded as equivalent in the chemistry literature, this alleged equivalence is actually  a mere resemblance that arises from the fact that the equations to be implemented in the closure scheme are the same in the case of only one trajectory. We also mention that  quantum-classical algorithms alternative to the Ehrenfest model are widely available. The most popular is the \emph{surface hopping method}, which however does not retain positivity of the quantum density matrix and thus invalidates Heisenberg's uncertainty principle.

In this paper, we propose to exploit the geometric variational structure of the new model \eqref{HybEq1}-\eqref{HybEq3} in order to make it amenable to  trajectory-based closures associated to the Lagrangian paths in the classical phase-space. Upon regularizing a suitable term in Hamilton's action principle, we will obtain a closure scheme that is formally the same as \eqref{MFeqs}, although the Hamiltonian $\widehat{H}_a$ is replaced by an effective Hamiltonian retaining correlation effects beyond the Ehrenfest theory. The resulting variational closure scheme will be illustrated after reviewing the formulation of the system in \eqref{HybEq1}-\eqref{HybEq3} and its geometric properties.

\section{Formulation of mixed quantum-classical models}

As mentioned in the Introduction, the  model \eqref{HybEq1}-\eqref{HybEq3} was formulated in \cite{GBTr22} by blending the symplectic geometry of Koopman's wavefunctions in classical mechanics with a gauge-invariance principle that arises from physical arguments. 

\paragraph{Koopman wavefunctions.}
As shown by Koopman in 1931, classical mechanics may be formulated as a unitary flow on the Hilbert space of square-integrable functions on phase-space. The main observation is that the {\it Koopman-von Neumann equation} (KvN) $i\hbar\partial_t \chi = \{i\hbar H,\chi\}$ yields the classical Liouville equation $\partial_tD=\{H,D\}$ for $D(\bz)=|\chi(\bz)|^2$. Importantly, the Liouvillian operator $\hat{L}_H=\{i\hbar H,\,\}$ is self-adjoint, thereby identifying a unitary evolution for $\chi$. 

Since both quantum and classical dynamics are written as unitary dynamics on Hilbert spaces, Sudarshan suggested to consider unitary evolution on the tensor-product space \cite{Sudarshan}. However, this turns out to be a difficult task and the first difficulty resides in the way  phases are treated in KvN theory. Indeed, writing $\chi=\sqrt{D}e^{iS/\hbar}$ gives $\de D/\de t=0$ and $\de S/\de t=0$ along $\dot{\bz}={\bf X}_H(\bz)$, so that the KvN phase evolution fails to reproduce the usual prescription arising from Hamilton-Jacobi theory, that is $\de S/\de t={\cal L}$, where ${\cal L}$ is the Lagrangian. This issue is readily addressed by modifying the Liouvillian $\hat{L}_H$ to include a phase term, so that the resulting {\it Koopman-van Hove equation} (KvH) reads
\beq\label{KvH}
i\hbar\partial_t\chi=\{i\hbar H,\chi\}-(p\partial_p H-H)\chi,
\eeq
where the  terms in parenthesis  evidently comprise the phase-space expression of the particle Lagrangian $\cal L$. In this case, the unitary time propagator is of the particular type $\chi(t)=(e^{-i\phi(t)/\hbar}\chi_0/\sqrt{\det\nabla \boldsymbol\upeta(t)})\circ\boldsymbol\upeta(t)^{-1}$, where $(\boldsymbol\upeta(t),e^{i\phi(t)/\hbar})$ is  an element of the infinite-dimensional group $\big\{(\boldsymbol\upeta,e^{i\phi/\hbar})\in \operatorname{Diff}(T^*Q)\,\circledS\,{\cal F}(T^*Q, S^1)\linebreak\big|\,\boldsymbol\upeta^*{\cal A}+\de\phi={\cal A}\big\}
$. 
Here, ${\cal F}(T^*Q,S^1)$ denotes the space of  functions on the phase space $T^*Q$ taking values in the unit circle, $\circledS$ denotes the semidirect-product, $\boldsymbol\upeta^*$ is the pullback, and ${\cal A}=p\de q$ is the Liouville one-form so that the canonical symplectic two-form reads $\omega=-\de{\cal A}$. Notice that the relation $\boldsymbol\upeta^*{\cal A}+\de\phi={\cal A}$ amounts to preservation of the Liouville one-form under the action of the semidirect-product group $\operatorname{Diff}(T^*Q)\,\circledS\,{\cal F}(T^*Q, S^1)$. Also, we observe that $\boldsymbol\upeta$ is a symplectic diffeomorphism, i.e. $\boldsymbol\upeta^*\omega=\omega$. The main advantage of the KvH equation \eqref{KvH}, first arisen in prequantization theory, is that it includes the correct prescription for the  phase evolution as well as reproducing the Liouville equation for the density. 

At this point, a first quantum-classical theory is obtained by starting with two classical systems and then quantizing one of them. This leads to the  {\it quantum-classical wave equation} (QCWE) for the hybrid wavefunction $\Upsilon(\bz,x)$ \cite{BoGBTr19}:
\beq\label{QCWE}
i\hbar\partial_t\Upsilon=\{i\hbar \widehat{H},\Upsilon\}-(p\partial_p \widehat{H}-\widehat{H})\Upsilon.
\eeq
As before, $\bz=(q,p)$ are classical  coordinates, $x$ is the quantum configuration coordinate, and $\widehat{H}(\bz)$ is an operator-valued function. Once again, the right-hand side of  equation \eqref{QCWE} identifies a self-adjoint operator which  leads to a unitary evolution of the hybrid wavefunction. The action principle $\delta\int_{t_1}^{t_2}\!\int\operatorname{Re}\big\langle\Upsilon\big|i\hbar\partial_t\Upsilon-\{i\hbar\widehat{H},\Upsilon\}+(p\partial_p\widehat{H}-\widehat{H})\Upsilon\big\rangle\,\de^2z\,\de t=0$ underlying \eqref{QCWE} identifies a Hamiltonian functional $h=\int\langle\widehat{\mathscr{D}}|\widehat{H}\rangle\,\de^2z$, where  ${\langle A|B\rangle=\operatorname{Tr}(A^\dagger B)}$ and $\widehat{\mathscr{D}}(\bz):=\Upsilon(\bz) \Upsilon^\dagger(\bz) + \partial_{p} (p \Upsilon(\bz)\Upsilon ^\dagger(\bz) )+{\rm i}\hbar \{\Upsilon(\bz) , \Upsilon^\dagger(\bz)\}$ is a measure-valued von Neumann operator.  Then, $\operatorname{Tr}\widehat{\mathscr{D}}$ is the classical density and $\int\widehat{\mathscr{D}}\,\de^2z$ is the quantum density matrix.

\paragraph{Phase symmetry in classical dynamics.}
While the QCWE has been studied extensively,  the unitary dynamics of hybrid wavefunctions does not appear  sufficient for  a consistent    theory. For example, the classical density $\operatorname{Tr}\widehat{\mathscr{D}}$ associated to \eqref{QCWE} is  generally sign-indefinite \cite{BoGBTr19}. As a further step, Sudarshan pointed out that classical phases, while  crucial to retain quantum-classical correlations, should eventually be made `unobservable'. We applied this idea by resorting to a `gauge principle' \cite{GBTr22}, that is by enforcing a symmetry under the group ${\cal F}(T^*Q, S^1)$ of phase transformations, in such a way that the latter are treated as a `gauge freedom'. For  this, one first needs to extract the classical phase from the hybrid wavefunction $\Upsilon$. This is accomplished by writing $\Upsilon(\bz,x)=\sqrt{D(\bz)}e^{iS(\bz)/\hbar}\psi(x;\bz)$, so that the last factor is a conditional quantum wavefunction and $S(\bz)$ is the classical phase. Replacing this factorization in the  action principle  underlying  the QCWE \eqref{QCWE} (see previous paragraph) yields $\delta\int_{t_1}^{t_2}\! L(D,S,\partial_tS,\psi,\partial_t \psi)\,\de t=0$, with
\beq
L=\!\int \!D\Big(\partial_tS-\operatorname{Re}\big\langle\psi\big|i\hbar\partial_t\psi-\{i\hbar\widehat{H},\psi\}+(p\partial_p\widehat{H}-\widehat{H})\psi+\nabla S\cdot{\bf X}_{\widehat{H}}\psi\big\rangle\Big)\,\de ^2z
\label{Tiziana2}
\eeq
and arbitrary variations $\delta D$, $\delta S$, and $\delta \psi$. Upon denoting $\langle\,,\rangle=\operatorname{Re}\langle\,|\,\rangle$, one realizes \cite{GBTr21} that replacing $\nabla S\to\boldsymbol{\cal A}+\langle\psi,i\hbar\nabla\psi\rangle$ makes the Hamiltonian functional $h(D,S,\psi)=\int D\big\langle\psi,(\widehat{H}-p\partial_p\widehat{H})\psi+\{i\hbar\widehat{H},\psi\}-\nabla S\cdot{\bf X}_{\widehat{H}}\psi\big\rangle\,\de ^2z$ gauge-invariant, i.e., invariant with respect to  $(D,S,\psi)\mapsto(D,S+\varphi,e^{-i\varphi/\hbar}\psi)$ for all  $\varphi(\bz,t)$. Here, 
$
{\boldsymbol{\cal A}=(p,0)}
$
 is the coordinate representation of the  one-form ${\cal A}={\boldsymbol{\cal A}\cdot\de\bz}=p\de q$.

In order to obtain an entire phase-invariant variational principle (not just a Hamiltonian functional), one  transforms the term $\int D\partial_t S\,\de ^2z$ in such a way to make $\nabla S$ appear explicitly and then  replaces $\nabla S\to\boldsymbol{\cal A}+\langle\psi,i\hbar\nabla\psi\rangle$. This was done in \cite{GBTr21} by noting that the equation $\partial_t D+\operatorname{div}(D\langle{\bf X}_{\widehat{H}}\rangle)=0$ resulting from the variations \eqref{Tiziana2} allows to use the dynamical relation $D(t)=\eta(t)_* D_0$, that is the density evolves by the push-forward of the initial condition $D_0$ by a time-dependent  Lagrangian path $\boldsymbol\eta(t)\in\operatorname{Diff}(T^*Q)$. Integration by parts with respect to time and  phase-space  leads to $\delta\int_{t_1}^{t_2}\!\int \!D\partial_t S\,\de^2z=-\delta\int_{t_1}^{t_2}\!\int\! D\nabla S\cdot\boldsymbol{\cal X}\,\de^2z$, where we have used $\partial_tD=-\operatorname{div}(D\boldsymbol{\cal X})$ and the vector field $\boldsymbol{\cal X}$ is such that $\dot{\boldsymbol\eta}=:\boldsymbol{\cal X}\circ\boldsymbol\eta$. Then, a phase-invariant action principle $\delta\int_{t_1}^{t_2}\! l(\boldsymbol{\cal X},D,\psi,\partial_t\psi)\,\de t=0$ is obtained upon replacing   \eqref{Tiziana2} by the Euler-Poincar\'e Lagrangian \cite{GBTr22}
\beq
l=\!\int \!D\Big(\boldsymbol{\cal X}\cdot\big(\boldsymbol{\cal A}+\langle\psi,i\hbar\nabla\psi\rangle\big)+\big\langle\psi,i\hbar\partial_t\psi-{\widehat{H}}\psi+i\hbar\big({\bX}_{\widehat{H}}-\langle{\bX}_{\widehat{H}}\rangle\big)\cdot\nabla\psi\big\rangle\Big)\,\de^2z.
\label{Tiziana2bis}
\eeq
Here,  the variations $\delta D$ and $\delta \boldsymbol{\cal X}$ are found to be constrained so that
\beq
\delta D=-\operatorname{div}(D\boldsymbol{\cal Y})
,\qquad\qquad\ 
\delta\boldsymbol{\cal X}=\partial_t\boldsymbol{\cal Y}+\boldsymbol{\cal X}\cdot\nabla\boldsymbol{\cal Y}-\boldsymbol{\cal Y}\cdot\nabla\boldsymbol{\cal X}
,
\eeq
where $\boldsymbol{\cal Y}=\delta\boldsymbol\eta\circ\boldsymbol\eta^{-1}$ is arbitrary.
Finally, one last convenient step consists in writing $\psi(t)=(U(t)\psi_0)\circ\boldsymbol\eta(t)^{-1}$, without loss of generality \cite{GBTr22}. Here, $U(t)=U(\bz,t)$ is a unitary operator on the quantum Hilbert space that is parameterized by phase-space coordinates. This step amounts to expressing the quantum unitary dynamics  in the  frame of Lagrangian classical paths. In this way,  the  Lagrangian \eqref{Tiziana2bis} is entirely expressed in terms of  $P=\psi\psi^\dagger$, i.e. it becomes {\it gauge-independent}.

\section{Geometry of quantum-classical dynamics}

The quantum-classical model \eqref{HybEq1}-\eqref{HybEq3} follows  from the variational principle associated to \eqref{Tiziana2bis}. We will now review the high points of its underlying geometry.

\paragraph{Euler-Poincar\'e variational principle.} Expressing the quantum evolution in the classical frame, or, equivalently, setting $\psi(t)=(U(t)\psi_0)\circ\boldsymbol\eta(t)^{-1}$ in  \eqref{Tiziana2bis}, leads to the  action principle $\delta\int_{t_1}^{t_2}\!\ell\,\de t=0$ for the following Lagrangian:
\beq
\ell(\boldsymbol{\cal X},D,\xi,{\cal P})=\int \!\big(D\boldsymbol{\cal A}\cdot\boldsymbol{\cal X}+\big\langle {\cal P},i\hbar\xi-{\widehat{H}}-i\hbar D^{-1}\{{\cal P},\widehat{H}\}\rangle\big)\,\de^2z
\,.
\label{Tiziana3}
\eeq
Here, $\langle\,,\rangle=\operatorname{Re}\langle\,|\,\rangle$,  ${{\cal P}=D\psi\psi^\dagger}$, and $\xi=(\dot{U}U^\dagger)\circ\boldsymbol\eta^{-1}$ is  skew-Hermitian,  so that 
\beq
\delta {\cal P} = [\Sigma, { \cal P}] - \operatorname{div}({\cal P} \boldsymbol{\cal Y})
,\qquad\qquad\ 
\delta \xi = \partial _t \Sigma + [ \Sigma , \xi  ] + \boldsymbol{\cal X}  \cdot\nabla  \Sigma   -  \boldsymbol{\cal Y}\cdot\nabla   \xi 
,
\eeq
where $\Sigma=(\delta{U}U^\dagger)\circ\boldsymbol\eta^{-1}$ is skew-Hermitian and arbitrary.
These variations arise by standard  Euler-Poincar\'e reduction from  Lagrangian to  Eulerian variables. Indeed,  Lagrangian trajectories play a crucial role in the variational problem associated to \eqref{Tiziana3}. In particular, if $\boldsymbol\eta(\bz_0,t)$ is the diffeomorphic Lagrangian path on phase-space and $U(\bz,t)$ is a unitary operator, we define the Eulerian quantities 
\beq
D:=\eta_*D_0
,\qquad\quad
{\cal P}:=\eta_*(U{\cal P}_0U^\dagger)
,\qquad\quad
\boldsymbol{\cal X}:=\dot{\boldsymbol\eta}\circ\boldsymbol\eta^{-1}
,\qquad\quad
\xi:=\dot{U}U^\dagger\circ\boldsymbol\eta^{-1}.
\label{LtoE}
\eeq
Then, taking the time derivative of the first two in \eqref{LtoE}  yields $\partial_t D+\operatorname{div}(D\boldsymbol{\cal X})=0$ and ${\partial_t{\cal P}+\operatorname{div}(\boldsymbol{\cal X}{\cal P})=[\xi,{\cal P}]}$, respectively. Furthermore, upon taking variations of \eqref{Tiziana3}, the action principle $\delta\int_{t_1}^{t_2}\ell\,\de t=0$ yields
\beq
\boldsymbol{{\cal X}}=\bX_{\textstyle\frac{\delta h}{
\delta D}}+\,\Big\langle \bX_{\textstyle\frac{\delta h}{\delta {\cal P}}}\Big\rangle
,\quad\ 
\Big[i\hbar  \xi-\frac{\delta h}{
\delta {\cal P}},{\cal P}\Big]=0,
\quad\,  \text{where}\quad
h=\!\int\!\langle D {\widehat{H}}+i\hbar \{{\cal P},\widehat{H}\}\rangle\de^2z
\label{EPeqns}
\eeq
and we have used ${\langle \widehat{A}\rangle:=\langle\psi,  \widehat{A}\psi\rangle}=D^{-1}\langle{\cal P}, \widehat{A}\rangle$. Then, after various manipulations we recover the system \eqref{HybEq1}-\eqref{HybEq3}. The purely quantum and classical cases are recovered by restricting to the cases $\bX_{\widehat{H}}=0$ and $\widehat{H}=H\boldsymbol{1}$, respectively \cite{GBTr22}. In addition, if one neglects the $\hbar-$terms in the Hamiltonian functional $h$, then the variational principle \eqref{Tiziana3} recovers the Ehrenfest model.

 Notice that the first two in \eqref{LtoE} indicate that the evolution of $D$ and ${\cal P}$  occurs on orbits of the semidirect-product group $\operatorname{Diff}(T^*Q)\circledS {\cal F}(T^*Q,{\cal U}(\mathscr{H}))$, where  ${\cal F}(T^*Q,{\cal U}(\mathscr{H}))$ denotes the space of phase-space functions taking values in the group ${\cal U}(\mathscr{H})$ of unitary operators on the quantum Hilbert space $\mathscr{H}$. In particular, these orbits are determined by the group action given by the composition of the standard  conjugation representation of ${\cal F}(T^*Q,{\cal U}(\mathscr{H}))$ and the pushforward action of $\operatorname{Diff}(T^*Q)$. The latter   diffeomorphism group comprises Lagrangian  paths on the classical phase-space.

\paragraph{Hamiltonian structure.} While the Hamiltonian structure of the model \eqref{HybEq1}-\eqref{HybEq3} is not necessary towards the development of the trajectory-based closure presented later, we quickly review it here as we are not aware of similar structures occurring elsewhere in continuum mechanics. Notice that the same Hamiltonian structure also applies to the Ehrenfest model \eqref{Ehrenfest}, which indeed is recovered by  neglecting the $\hbar-$terms in the Hamiltonian functional $h$ in \eqref{EPeqns}.  Thus, all the considerations in this discussion apply equivalently to the Ehrenfest model \eqref{Ehrenfest}.

First, we observe that the variable $D$ may be written as $D=\operatorname{Tr}{\cal P}$ so that the Euler-Poincar\'e Lagrangian \eqref{Tiziana3} is expressed entirely in terms of the variables $(\boldsymbol{\cal X},\xi,P)$. Going through the same steps as above leads to rewriting \eqref{EPeqns} as
$
\boldsymbol{{\cal X}}=\langle \bX_{{\delta h}/{\delta {\cal P}}}\rangle
$ and $
{[i\hbar  \xi-{\delta h}/{
\delta {\cal P}},{\cal P}]=0}
$, where ${
h=\int\langle  {\widehat{H}}\operatorname{Tr}{\cal P}+i\hbar \{{\cal P},\widehat{H}\}\rangle\,\de^2z}$. Then, the Hamiltonian equation 
\beq
i\hbar\frac{\partial\cal P}{\partial t}+i\hbar\operatorname{div}\!\Big({\cal P}\Big\langle \bX_{\textstyle\frac{\delta h}{\delta {\cal P}}}\Big\rangle\Big)=\Big[\frac{\delta h}{\delta {\cal P}},{\cal P}\Big]
\label{HamEqn}
\eeq
leads directly to the following bracket structure via the usual relation $\dot{f}=\{\!\!\{f,h\}\!\!\}$:
\begin{equation}\label{bracket_candidate_rho}
\{\!\!\{f,h\}\!\!\}(\mathcal{P})=\int \!\frac1{\operatorname{Tr}\mathcal{P}  }\bigg(\mathcal{P}  : \bigg\{\frac{\delta f}{\delta \mathcal{P}},\frac{\delta h}{\delta \mathcal{P}}\bigg\}: \mathcal{P}   \bigg)\de^2z 
 -
\int\!\left\langle \mathcal{P}  ,\frac{i}\hbar\!\left[\frac{\delta f}{\delta \mathcal{P}},\frac{\delta h}{\delta \mathcal{P}}\right] \right\rangle\de^2z,
\end{equation} 
where we have introduced the convenient notation ${A:B}=\operatorname{Tr}(AB)$. The proof that \eqref{bracket_candidate_rho} is Poisson involves a combination of results in Lagrangian and Poisson reduction  \cite{GBTr21}. The first term in \eqref{bracket_candidate_rho} is related to the   Lagrangian classical paths.

Equation \eqref{HamEqn} easily leads to  characterizing the Casimir invariant $
C_1=\operatorname{Tr}\!\int\! D\Phi({\cal P}/D)\,\de^2z
$ for any matrix analytic function $\Phi$. Also, upon writing ${\cal P}=D\psi\psi^\dagger$, one finds the quantum-classical Poincar\'e integral invariant
\[
\frac{\de}{\de t}\oint_{\boldsymbol{c}(t)}\big(p\de q+\langle\psi,i\hbar\de\psi\rangle\big)=0
\]
for any loop $\boldsymbol{c}(t)=\boldsymbol\eta(\boldsymbol{c}_0,t)$ in phase-space. Here, we notice the important role of the Berry connection $\langle\psi,-i\hbar\de\psi\rangle$. By Stokes theorem, the above relation also allows to identify a Lie-transported quantum-classical two-form on $T^*Q$, that is
\[
\Omega(t)=\eta_*\Omega(0)
, 
\qquad\text{ with }\qquad
\Omega(t):=\omega+\hbar\operatorname{Im}{\langle\de\psi(t)|\wedge\de\psi(t)\rangle}
,
\]
so that $\Omega(t)$ remains symplectic in time if it is so initially. As a result, if $\operatorname{dim}Q=n$, one finds the additional class of Casimirs
$
C_2=\int\! D\Lambda\big(D^{-1}\Omega^{\wedge n}\big)\de^{2n}z
$,
where $\Omega^{\wedge n}=\Omega\wedge\dots\wedge\Omega$ ($n$ times) is a volume form  and $\Lambda$ is  any scalar function of one variable.  These Casimirs may be used to construct quantum-classical extensions of  Gibbs/von Neumann entropies \cite{GBTr22}. If $\operatorname{dim}Q=1$, then $\Omega=(1+\hbar\operatorname{Im}\{\psi^\dagger,\psi\})\omega$.

\paragraph{Quantum-classical von Neumann operator.} We observe that the Hamiltonian energy functional $h$ in \eqref{EPeqns} is not simply given by the usual average of the  Hamiltonian operator $\widehat{H}$. Indeed, the $\hbar-$term seems to play a crucial role in taking the model \eqref{HybEq1}-\eqref{HybEq3} beyond simple Ehrenfest dynamics. As discussed in \cite{GBTr22}, this suggests that the quantum-classical correlations trigger extra energy terms that are not usually considered. Alternatively, one may insist that the total energy must be given by an average of $\widehat{H}$. Following this route leads to rewriting the last in \eqref{EPeqns} as $h=\operatorname{Tr}\int\!\widehat{\cal D}\widehat{H}\,\de^2 z$, where 
\[
\widehat{\cal D}=D{P}+\frac{\hbar}2\operatorname{div}(D\widehat{\boldsymbol\Gamma})
=
D{P}+\frac{i\hbar}2\operatorname{div}(D[P,\bX_P])
\] 
is a measure-valued von Neumann operator and we recall $P=\psi\psi^\dagger$. Then, classical and quantum densities are  given  by taking the   trace and integral of $\widehat{\cal D}$, respectively. Unlike the quantum density operator,  the hybrid operator $\widehat{\cal D}$ is not sign-definite. Remarkably, however, $\widehat{\cal D}$ enjoys the  equivariance properties
\[
\widehat{\cal D}(\boldsymbol\upeta_*D,\boldsymbol\upeta_*P)=\boldsymbol\upeta_*\widehat{\cal D}(D,P),
\qquad\text{and} \qquad
\widehat{\cal D}(D,\mathscr{U}P\mathscr{U}^\dagger)=\mathscr{U}\widehat{\cal D}(D,P)\mathscr{U}^\dagger,
\] 
where $\boldsymbol\upeta$ is a symplectic diffeomorphism on $T^*Q$ and $\mathscr{U}\in{\cal U}(\mathscr{H})$.
These two properties ensure the following dynamics in the classical and quantum sector \cite{GBTr21}:
\[
\frac{\partial D}{\partial t}=\operatorname{Tr}\{\widehat{H},\widehat{\cal D}\}
\,,\qquad\qquad
i\hbar\frac{\de\hat\rho}{\de t}=\int[\widehat{H},\widehat{\cal D}]\,\de^2 z
\,.
\]
For example, the first can be verified directly upon writing ${\boldsymbol{\cal X}} =D^{-1}\langle\widehat{\cal D}, \bX_ {\widehat{H}}\rangle+\hbar D^{-1\!}\operatorname{div}\!\big(D\operatorname{Tr}( \bX_ {\widehat{H}}\wedge\widehat{\boldsymbol\Gamma})\big)$, where  $
( \bX_ {\widehat{H}}\wedge\widehat{\boldsymbol\Gamma})^{jk}:=\big( X_ {\widehat{H}\,}^j{\widehat{\Gamma}}^k-{\widehat{\Gamma}}^j X_ {\widehat{H}}^k\big)/2$ identifies a bivector. 

\section{Trajectory-based closure}
As discussed in the Introduction, the quantum-classical model \eqref{HybEq1}-\eqref{HybEq3} does not immediately allow for the application of the trajectory-based closure typically adopted for the Ehrenfest equations. This is due to the  appearance of several gradients in the expressions \eqref{HybEq2} and \eqref{HybEq3}. A similar situation also occurs in quantum hydrodynamics, thereby preventing the existence of particle solutions in Bohmian mechanics \cite{FoHoTr19}. In the latter case, a regularization technique was recently introduced to allow the standard application of trajectory-based closures.  Unlike common regularizations, this particular one was introduced at the level of the variational principle and its successful implementation was presented in \cite{HoRaTr21}. The resulting closure  arises from a sampling process at the level of the classical Lagrangian paths. This section exploits this approach in such a way to formulate a trajectory-based closure of the quantum-classical model \eqref{HybEq1}-\eqref{HybEq3}. In particular, we will devise a computational method that inherits  basic conservation laws, such as energy and total probability, and retains decoherence effects beyond the standard Ehrenfest model.

\paragraph{Variational regularization.}
The present method arises from the observation that the singular solution ansatz ${\cal P}(\bz,t)=\sum_{a=1}^N w_a \rho_a(t) \delta(\bz-\bzeta_a(t))$ is prevented by  the last term in the Hamiltonian $h$ in \eqref{EPeqns}. Thus, if a regularization needs to be introduced at the variational level, it has to be introduced in that term. Here, we will replace the Lagrangian \eqref{Tiziana3} by  the {\it regularized Lagrangian}
\[
\bar\ell=\int\!\Big(D\bA\cdot\boldsymbol{\cal X}+\langle{\cal P},i\hbar\xi-\widehat{H}\big\rangle-\frac12\big\langle\bar{\cal P},i\hbar \bar{D}^{-1}\big[\nabla\bar{\cal P},\bX_{\widehat{H}}\big]\big\rangle\Big)\de^6 z
\,,
\]
where the commutator arises from conveniently  projecting $i\{{\cal P},\widehat{H}\}$ on its Hermitian part, and we have introduced the regularized quantities
\[
\bar{D}=\int\! K_\alpha(\bz-\bz') D(\bz')\,\de^2 z'
\,,\qquad\qquad
\bar{\cal P}=\int\! K_\alpha(\bz-\bz') {\cal P}(\bz')\,\de^2 z'.
\]
The \emph{mollifier} $K_\alpha$ is chosen as a smooth convolution kernel that is invariant under phase-space translations and tends to the delta function as $\alpha\to0$, that is the limit in which one recovers the original model. For example, $K_\alpha$ may be a Gaussian kernel with variance $\alpha$, although here we will keep it general. With this regularization,  one allows to consider the singular solution ansatz 
\beq
D=\sum_{a=1}^N w_a\delta(\bz-\bzeta_a)
\,,\qquad\qquad
{\cal P}=\sum_{a=1}^Nw_a\rho_a\delta(\bz-\bzeta_a).
\label{ansatz}
\eeq
For example, the first in \eqref{HybEq1} now leads to 
$
\dot\bzeta_a=\boldsymbol{\cal X}_a
$ with ${\boldsymbol{\cal X}_a:=\boldsymbol{\cal X}(\bzeta_a)}$.
Similarly, the equation  $i\hbar\partial_t{\cal P}+i\hbar\operatorname{div}({\cal P}\boldsymbol{\cal X})=[\xi,{\cal P}]$ (see previous section)
leads to
$ \dot\rho_a=[\xi_a,\rho_a]
$ with 
$
\xi_a:=\xi(\bzeta_a)$. Here, we will set $\rho_a=\psi_a\psi^\dagger_a$ so that $\dot\psi_a=\xi_a\psi_a$. We remark that, as in the case of the Ehrenfest model,  the trajectories $\bzeta_a(t)$ in \eqref{ansatz} are \emph{not} physical particles, but rather arise from a sampling process of the Lagrangian classical paths underlying the Eulerian \makebox{action principle associated to \eqref{Tiziana3}.}

\paragraph{Trajectory equations.} At this stage, we are ready to replace the ansatz \eqref{ansatz} in the regularized Lagrangian, thereby obtaining the   finite-dimensional Lagrangian
\beq
L(\{\bzeta_a\},\{\xi_a\},\{\rho_a\})=\sum_aw_a\bigg(
p_a\dot{q}_a+\bigg\langle\rho_a,i\hbar\xi_a-{\widehat{H}}_a-i\hbar
\sum_{b}w_b  [{\rho}_b,{\cal I}_{ab}]\bigg\rangle\bigg)
.
\label{KoopLagr}
\eeq
Here, $\delta \xi_a = \dot{\Sigma}_a + [ \Sigma_a , \xi_a  ]$, with $\Sigma_a$ arbitrary, and we have denoted
\[
{\cal I}_{ab}:=\frac12\int\frac{K_a\{K_b,\widehat{H}\}}{\sum_c w_c K_c}\,\de^2z
\,,\qquad\text{ and }\qquad
K_s(\bz,t):=K(\bz-\bzeta_s(t))
\,.
\]
Once again, we observe that if  the $\hbar-$terms are neglected in \eqref{KoopLagr}, then the associated variational principle recovers the closure equations \eqref{MFeqs} associated to the Ehrenfest model. Instead, in the general case each  trajectory is directly coupled to all the others via the $\hbar-$term.  The equations of motion read
\beq
\dot{q}_a=w_a^{-1}{\partial_{p_a\!} h}
,\qquad\ \ 
\dot{p}_a=-w_a^{-1}{\partial_{q_a\!} h}
,\qquad\ \ 
i\hbar{\dot{\rho}_a}=w_a^{-1}[{\partial_{\rho_a\!} h},\rho_a]
,
\label{MFeqs2}
\eeq
where 
\[
h=\sum_aw_a
\bigg\langle\rho_a,{\widehat{H}}_a+i\hbar
\sum_{b}w_b  [{\rho}_b,{\cal I}_{ab}]\bigg\rangle
,\qquad\quad
\partial_{\rho_{a\!}} h={\widehat{H}}_a+i\hbar
\sum_{b}w_b  [{\rho}_b,{\cal I}_{ab}-{\cal I}_{ba}]
.
\]
In analogy to the discussion in the previous section, we can rearrange the Hamiltonian $h$ above as $h=\operatorname{Tr}\int\widehat{\cal D}\widehat{H}\,\de^2z$ with the hybrid von Neumann operator
\[
\widehat{\cal D}(\bz,t)= \sum_aw_a\hat{\rho}_a(t)\delta(\bz-\bzeta_a(t))+{i\hbar}\sum_{a,b}w_aw_b  {\cal J}_{ab}(\bz,t)\big[\hat{\rho}_a(t),\hat{\rho}_b(t)\big],
\]
where
\[
{\cal J}_{ab}:=\frac14
\bigg(\bigg\{K_a,\frac{K_b}{\sum_c w_c K_c}\bigg\}
-\bigg\{K_b,\frac{K_a}{\sum_c w_c K_c}\bigg\}\bigg).
\]
The implementation of this closure scheme is currently underway. We observe that the canonical Hamiltonian structure underlying this scheme may pave the way to the application of symplectic integration techniques for the long-time   simulation of fully nonlinear processes.


%
%
%
%

\end{document}